# Nine Recommendations for Decision Aid Implementation from the Clinician Perspective

Anshu Ankolekar[1], Ben G.L. Vanneste[1], Esther Bloemen-van Gurp[2,6], Joep van Roermund[3], Adriana Berlanga[4], Cheryl Roumen[1], Evert van Limbergen[1], Ludy Lutgens[1], Tom Marcelissen[3], Philippe Lambin[5], Andre Dekker[1], Rianne Fijten[1]

[1]Department of Radiation Oncology (MAASTRO), GROW School for Oncology and Developmental Biology, Maastricht University Medical Centre+, Maastricht, The Netherlands

[2]Fontys University of Applied Sciences, Eindhoven, The Netherlands

[3]Department of Urology, Maastricht University Medical Centre+, Maastricht, The Netherlands

[4]Maastricht University, Maastricht, The Netherlands

[5]The D-Lab, Department of Precision Medicine, GROW - School for Oncology, Maastricht University Medical Centre+, Maastricht University, Maastricht, The Netherlands

[6]Zuyd University of Applied Sciences, Heerlen, The Netherlands

**Abstract**

**Background**

Shared decision-making (SDM) aims to empower patients to take an active role in their treatment choices, supported by clinicians and patient decision aids (PDAs). The purpose of this study is to explore barriers and possible facilitators to SDM and a PDA in the prostate cancer trajectory. In the process we identify possible actions that organizations and individuals can take to support implementation in practice.

**Methods**

We use the Ottawa Model of Research Use as a framework to determine the barriers and facilitators to SDM and PDAs from the perspective of clinicians. Semi-structured interviews were conducted with urologists (n=4), radiation oncologists (n=3), and oncology nurses (n=2), focusing on the current decision-making process experienced by these stakeholders. Questions included their attitudes towards SDM and PDAs, barriers to implementation and possible strategies to overcome them.

**Results**

Time pressure and patient characteristics were cited as major barriers by 55% of the clinicians we interviewed. Structural factors such as external quotas for certain treatment procedures were also considered as barriers by 44% of the clinicians. Facilitating factors involved organizational changes to embed PDAs in the treatment trajectory, training in using PDAs as a tool for SDM, and clinician motivation by disseminating positive clinical outcomes. Our findings also suggest a role for external stakeholders such as healthcare insurers in creating economic incentives to facilitate implementation.

**Conclusion**

Our findings highlight the importance of a multi-faceted implementation strategy to support SDM. While clinician motivation and patient activation are essential, structural/economic barriers may hamper implementation. Action must also be taken at the administrative and policy levels to foster a collaborative environment for SDM and, in the process, for PDAs.

**Background**

Shared decision-making (SDM) is the process of collaboration between clinicians and patients to come to a treatment decision based on the best available clinical information and the patient's personal preferences (1). Making treatment choices in oncology is becoming increasingly complex; patients must not only absorb a large amount of clinical information but also weigh multiple treatment options in terms of benefits and harms while confronting the emotional and psychological aspects of their disease (2). Lack of awareness about different treatment options can result in patients making choices that are not aligned with their values (3). Such decisions often lead to decisional regret (4, 5). Patient decision aids (PDAs) are tools that help inform patients about their treatment options and in some cases help them clarify their values. PDAs are often used to facilitate SDM. A growing body of evidence suggests that patients who use PDAs are better informed about their condition and preferences, have better risk perceptions, and have lower decisional conflict compared to patients who receive usual care (3).

In the Netherlands, policymakers at various levels (government, healthcare insurers, and patient advocacy groups) are actively facilitating an increased role for patients in the management of their own health through initiatives such as the Dutch Decision Aids Implementation Program (D-DAP) (6). These initiatives are yet to translate into increased patient involvement on the whole and challenges remain in integrating PDAs in clinical practice (7). These include poor design of the tools themselves and a lack of usability (8). Clinicians may lack awareness of PDAs (9) or may believe that they are of limited value in practice (10). Integration into existing workflows presents a structural barrier (11), along with an ongoing need to keep PDAs up to date in order to keep pace with the rapidly changing landscape of oncological research. As yet, there has been little discussion on what organizational changes can be made to support the use of PDAs in clinical practice.

Furthermore, research on implementation tends to conflate SDM and PDAs (12). For instance, a systematic review on barriers and facilitators to SDM defines SDM both as the process of clinician-patient engagement as well as an intervention such as a PDA (13). However, providing a PDA does not guarantee that SDM will take place, and a PDA implemented in an environment that is not conducive to SDM may not show an improvement in outcomes (14). Therefore, when introducing a PDA into a clinical context it is worthwhile to consider the barriers and facilitators not only to the tool itself but also to SDM in the environment the tool will be implemented in. With this knowledge it may be possible to derive an implementation strategy that uses the PDA to support SDM.

This study was conducted as a precursor to implementing a web-based prostate cancer PDA developed at our clinic (15). We began with the premise that PDAs may be a useful tool for SDM to the extent that SDM can be successfully implemented in the clinical environment. Any implementation strategy should therefore consider both the barriers and facilitators to SDM and the resulting implications for the PDA. We explore these factors in parallel from the clinician perspective.

**Methods**

We used the Ottawa model of research use (OMRU) as a framework for our data collection (16). The OMRU outlines three factors that must be monitored when implementing a new health intervention (see Fig. 1): the intervention itself (in this context, the PDA), the readiness of the potential adopters (practitioners/patients), and factors in the practice environment, such as culture, workflows, and resources, that may be barriers or supports.

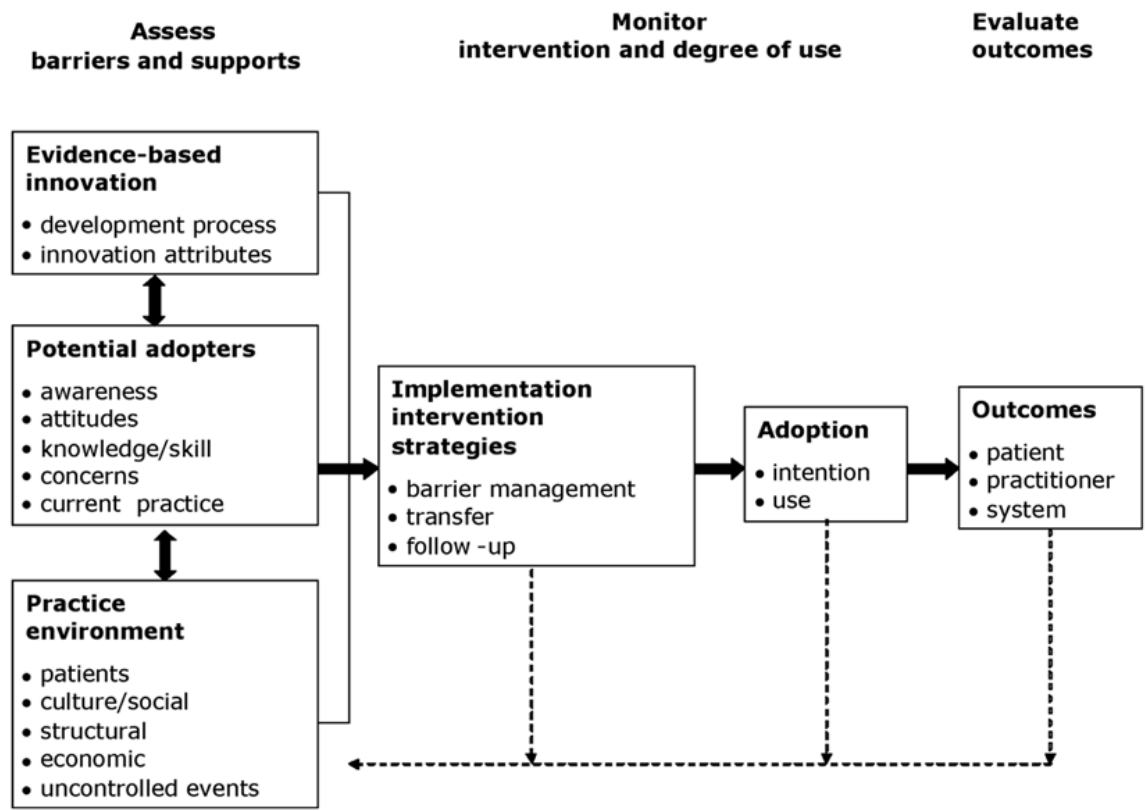

*Figure 1: Ottawa model of research use (17)*

These factors were studied by interviewing nine clinicians specialized in prostate cancer (four urologists, three radiation oncologists, and two oncology nurses) within our clinic in a semi-structured interview format. The clinicians ranged in experience commensurate with their age. Characteristics of the participants are displayed in Fig. 2. Interview questions were based on the following components of the OMRU: current treatment process, awareness of SDM principles, attitudes towards SDM and PDAs, beliefs about external/organizational factors that might inhibit the application of SDM and use of PDAs, and lastly actions that can be taken within the organization to foster PDA uptake. Interview questions covered their perceptions of SDM and PDAs in general as well as of the prostate cancer PDA developed previously. Participants were sent a link to the PDA two weeks before the interview to familiarize themselves with it.

Interviews lasted between 30 to 60 minutes and were taped using a voice recorder. The recordings were then transcribed fully and emailed back to the participants for their review and approval. After approval, each transcript was analyzed by means of open coding in which text fragments were assigned codes summarizing their meaning. Subsequently, to identify relationships between the open codes, we performed axial coding by grouping the open codes according to their underlying themes. To improve the validity of the results a sample of the interviews were coded separately by two different researchers and compared to reconcile any differences.

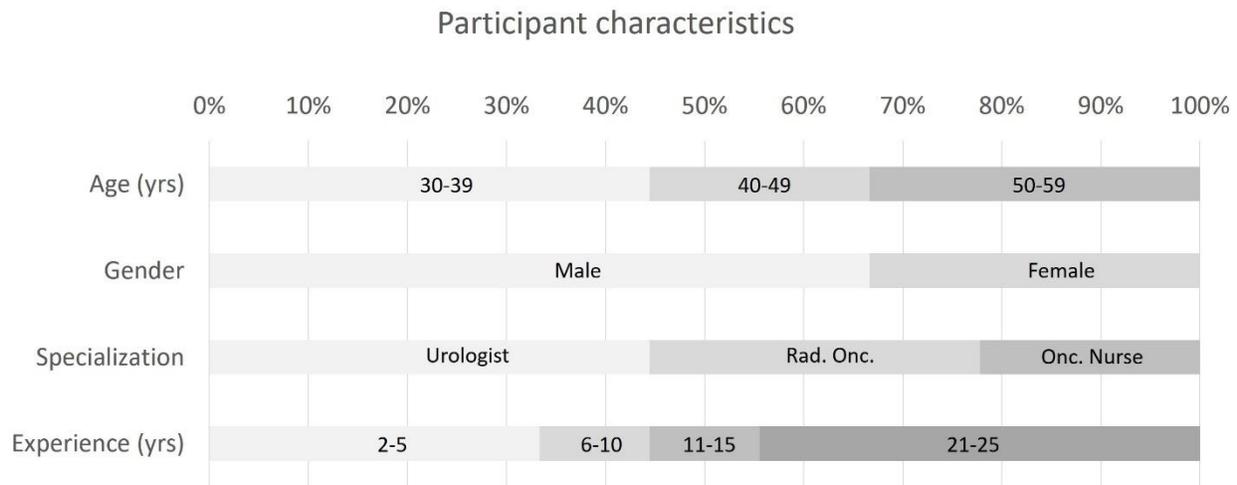

*Figure 2: Participant demographics*

## Results

Our analysis resulted in a list of barriers and facilitators to SDM at the individual, organizational and external level, as well as attributes of PDAs that could hinder or support to their implementation. The full list is summarized in Table 1.

**Barriers and facilitators to SDM**

*Potential adopters*

Our results revealed an overall positive attitude towards SDM among clinicians, with most expressing that following a patient's preference is ideal and that SDM should be more prevalent in clinical practice than it is currently. They mentioned several factors that may hinder SDM in practice, namely time pressure/efficiency, the nature of SDM itself, patient characteristics, and clinician characteristics.

More than half (five out of nine) clinicians mentioned time pressure as the biggest barrier to SDM. Some perceived it to be an inefficient process to engage in dialogue with the patient about his preferences. Two clinicians noted that it is easier to decide the treatment option themselves based on their clinical experience. Time constraints were a more significant barrier for urologists, with radiation oncologists reporting that they have enough time for the patients. Three urologists mentioned that their consultations average 10-15 minutes. Radiation oncologists said that they have around 45-60 minutes to discuss treatment details, and one nurse reported having 30-45 minutes for each patient and more time if necessary.

> 'I think that [SDM] is much less efficient, it is very time-consuming. [...] It is far easier to just say "This is [the best option]".' (Radiation oncologist 2)

Five clinician expressed doubts about SDM as a paradigm, highlighting the burden that SDM may place on the patient by making him responsible for treatment choices. Patients can get overwhelmed by the volume of clinical information needed to make treatment decisions. They may also lack complete or balanced information, which makes it more difficult for them to make an informed choice. The need to make a treatment choice may also induce stress and confusion.

> '[SDM] asks a lot of the patient. And that also means where things go wrong, which happens sometimes, of course, you also have those people who have been irradiated and who [regret it]. They say "I should have chosen operation." [...] We now place the choice with the patient and with it the responsibility more or less is on him.' (Urologist 3)

Finally, certain clinician characteristics were thought to influence the scope for SDM. Expertise was one characteristic; clinicians are more knowledgeable about their own specialization and may have less experience with other treatments. For instance, urologists are able to provide information on surgery in greater detail than on radiotherapy in which they have less direct expertise. However, they are the lead clinicians overseeing most of the treatment trajectory and are responsible for having the decisional talk with the patient. Four clinicians found this a significant barrier to SDM because incomplete information may prevent the patient from making a balanced choice. Another characteristic mentioned by one participant was the clinician's age, with the perception being that younger clinicians are more open to SDM while it requires a mentality change for their more senior peers.

> 'The age-old discussion is that people preach to the converted. [...] I would say the patient has to have surgery, a radiation oncologist would say he has to have radiotherapy.' (Urologist 1)

The biggest facilitator at the individual level was the motivation of the clinician leading the treatment process (in this case, the urologist). Clinicians reported that they would be more open to SDM if they were convinced that it leads to better patient outcomes and does not lengthen the decision-making process.

> '[They] must also see the benefit, that patients are more satisfied, that they get more insight into their disease, that they ultimately have a better quality of life. The person implementing or actually offering it to patients must be convinced of that.' (Radiation oncologist 1)

*Practice environment*

Participants cited two main barriers in the clinic environment: patient characteristics and structural factors. Patient characteristics included the patient's age, health literacy and willingness to participate in SDM. Three clinicians mentioned that the advanced age of the typical prostate cancer patient makes it challenging to engage the patient in the decision-making process. Furthermore, many older patients are used to a paternalistic model of decision-making in which the clinician advises the patient on the best treatment. The wide variety in patient backgrounds, some being more educated than others and better able to understand clinical information, presented another challenge according to one clinician. Finally, some patients are willing to take an active role while others prefer the clinician to take the lead, and even when decisions are made jointly, a considerable portion of patients are highly influenced by the clinician's opinion.

> 'You have to imagine that there are different [types] of patients. From highly educated to minimally educated and that involves a totally different manner of communication.' (Urologist 1)

> *'The patients, particularly the elderly patients [...], are more used to a paternalistic model, they believe [...] that the doctor knows what is best for them.' (Radiation oncologist 2)*

Four clinicians cited structural factors that may impede SDM. First, in the current treatment trajectory the urologist is the first point of contact for the patient and oversees most of the process, from initial diagnosis to the final decision. The urologist carries out the decisional talk with the patient and presents all relevant treatment options. If radiotherapy is chosen then the patient is referred to the radiation oncologist for further treatment, however many patients tend to choose surgery. Second, government-set quotas on the required number of surgeries to be performed by hospitals may incentivize treatment choices to be steered in that direction.

> *'There are guidelines in the Netherlands on prostatectomy. To be able to do such a prostate surgery you must achieve a minimum number per year.' (Radiation oncologist 3)*

> *'We also have to get a certain number of operations and that is also from the government.' (Urologist 4)*

Clinicians suggested several actions that would facilitate SDM implementation, beginning with making it an organizational priority. Creating the time and space for it, both literally in terms of more contact time with the patient and physical set-ups such as joint consultations, and logistically such as through training and addressing external factors such as treatment quotas. Clinicians suggested training to acquaint their peers with the principles of SDM and how to elicit participation from the patient. Some clinicians valued an e-Learning course about the basics of practicing SDM, others suggested face-to-face demonstrations so that clinicians could inform and educate their patients more effectively.

Joint consultations were proposed to reduce information asymmetry between the different specializations. If radiotherapists and urologists were to participate in consultations together with the patient this could improve the decision-making process by giving all parties a more balanced perspective of the options available and patient's preferences. Despite the potential benefits, most clinicians highlighted the organizational challenges this level of care coordination may present.

> *'An online [course] is the easiest as you can follow it in your own time […] the so-called e-Learning.' (Radiation oncologist 3)*

> *'It would be great to have an oncology center for example, where you could discuss everything with another specialist [...] but that is also difficult.' (Urologist 4)*

> *'The [hospital] must be ready [for] the extra logistical challenges […] to have two doctors in place of one, and preferably at the same time or at the same clinic.' (Radiation oncologist 3)*

**Barriers and facilitators to PDAs**

In light of existing time pressures, clinicians said they would be less likely to use the PDA if it was too complicated to understand or apply. This included factors such as the time taken to navigate through the PDA, fill in information such as patient details, and discuss the outputs in the consultation with the patient. Another potential barrier was accessibility in the case of a web-based PDA; patients, particularly the elderly, may not be familiar enough with PCs/tablets to use the PDA.

> *'If we have to log in […] it will cost us a lot of time and energy. It should therefore remain simple.' (Urologist 4)*

> *'If the decision aid is too difficult or too complicated then it is not usable. So you have to adjust it to the mainstream population of course." (Radiation oncologist 1)*

> *'So people have to have access to a computer, access to the internet. If there are people who really don't understand or who can't read or who can't use a computer or don't have internet access, yes then it is difficult.' (Urologist 2)*

Numerous facilitators were proposed, both on the PDA-level as well as in the development process. Three radiation oncologists and two urologists cited the critical role of the urologist in implementing a PDA as urologists are the first point of contact for the patient after referral and oversee a large part of the treatment trajectory. Motivating the urologist was seen as a stimulating factor, as the clinician who offers the PDA must be convinced of its added value. This can be enhanced by showing positive results associated with the use of PDAs, for example that patients may face less decisional regret and accept the side effects and complications more readily if he has opted for treatment himself.

At a structural level, embedding the PDA in the patient pathway in a standardized manner was seen as an ideal way to make the informing process more efficient. Oncology nurses were named as the most effective channel through which the patients could be introduced to the PDA and guided through the process of applying it, since they were regarded as more approachable for the patients. Involving additional personnel in this manner was thought to save time during consultations. The present role of oncology nurses is to provide the patient with practical and logistical information about their treatment options. These include the procedures for the different treatments and the implications in terms of hospital visits and so forth. We did not find evidence that nurses engage the patient in actual deliberation or decision-making; this function appears to be performed solely by the clinicians. Finally, two clinicians mentioned compensation for the use of PDA as a facilitating factor.

> *'All the diagnostics and the workup up to and including the diagnosis, plus discussing the tumor board decision are all with the urologist himself. So he is the key player; if he does not use it or does not recommend it, nothing happens.' (Radiation oncologist 3)*

> *'[Implementation] is only possible if all clinicians are obliged to use [the PDA]. But that will never happen of course. Unless instead, you get compensation from the health insurance company.' (Radiation oncologist 1)*

*Table 1: Summary of barriers and facilitators to SDM and PDA implementation from clinicians' perspective and the number of clinicians who mentioned each factor*

| Level | Barrier | N | Facilitator | N |
|---|---|---|---|---|
| **Potential adopters** | | | | |
| Attitudes towards SDM | • Burden on patient<br>• Takes too much time | 5<br>5 | • Better patient outcomes<br>• Clinician motivation | 3<br>3 |
| Knowledge/skill | • Expertise in specialization | 3 | | |
| **Practice environment** | | | | |
| Patient characteristics | • Age<br>• Health literacy<br>• Willingness to participate | 3<br>3<br>2 | | |
| Cultural/social | • Paternalistic attitudes | 2 | | |
| Structural | • Not enough time<br>• Volume quotas | 4<br>4 | • Training<br>• Organizational priority<br>• More contact time with patient | 5<br>4<br>2 |
| Economic | | | • Compensation | 2 |
| **PDA** | | | | |
| Development process | | | • Clinician involvement | 3 |
| Attributes | • Requires computer skills<br>• Time-consuming<br>• Biased information<br>• Complicated | 3<br>2<br>2<br>2 | • Balanced information<br>• Embedded in workflow<br>• Personalized<br>• Interactive | <br>3<br>2<br>2 |

**Discussion**

Implementing a new health intervention involves action on multiple levels, centering on the intervention itself, the people who will use it, and the environment in which it will be introduced. In this study, we interviewed clinicians to understand their perspective on the factors that can hinder or support SDM implementation, and what actions may support the use of a PDA. Overall, our findings corroborate existing evidence that time pressure and patient characteristics are perceived as the biggest barriers to SDM (13). Our study also sheds light on barriers that have received little attention in the literature, such as certain structural and economic incentives that may affect PDA uptake and the scope for SDM. In analyzing our findings in the context of existing literature, we propose nine recommendations for PDA implementation. Our list is by no means exhaustive but address some of the more prominent issues in SDM implementation and suggests actions that may increase the likelihood of implementation success.

*Recommendation 1: Make SDM and PDA implementation an organizational priority*

Current literature on SDM tends to focus on the interaction between clinician and patient, however organizational factors can heavily influence the implementation of PDAs and other health interventions. These factors include, among others, the organizational culture, available resources, and clinical workflows (18). Our participants emphasized the importance of making SDM initiatives an organizational priority through strategies such as coordination across specializations (e.g. joint consultations). This theme is echoed in several studies, namely that actions taken on multiple levels are more effective than those that target a single level (19, 20). Strong leadership plays a key role in coordinating implementation efforts across the organization and improving quality of care (21). Simultaneously, empowering clinicians by giving them a degree of autonomy in SDM agendas may make implementation efforts more effective (22). Examples may include inviting clinician participation and feedback in goal setting and developing SDM training activities.

*Recommendation 2: Create ownership*

Clinician motivation was considered one of the biggest facilitators by our study participants. This is also the top-most factor mentioned in a systematic review of barriers and facilitators (13). Prior studies have shown that uptake of innovations in general, and PDAs specifically, is enhanced when end users are involved throughout the development process (23-27). Involving clinicians in PDA development and evaluation may help create trust and ownership in the PDA and make it more likely that these clinicians will use the PDA in their practice. If the PDA is to be implemented in several clinics or geographical areas, ownership could be created by providing customization options (28), such as the possibility for other clinics to add their own logo or informational videos made with their own clinicians to the PDA.

*Recommendation 3: Make the PDA interactive and/or personalized*

SDM is, by definition, an interactive process between clinician and patient. The former provides clinical information and the latter shares their preferences and what matters to them so that both can make a treatment decision together. Eliciting patient preferences was cited as a challenge for clinicians in our study and other studies (29). In particular, elderly cancer patients who are used to a paternalistic decision-making process are more prone to following the clinician's advice. An interactive and/or personalized PDA may prompt patients to engage in the SDM process more actively and provide a platform for sharing personal preferences and values. One solution is to include an interactive element in the PDA, such as filling in values clarification questions which stimulate the patient to consider what matters to him. Patients may not always know their preferences beforehand, and the act of weighing treatment options and discussing the possibilities with a clinician can help preferences emerge (30). These results can then be summarized and presented back to the patient for further reflection and taken into the consultation as an input for the decisional talk.

*Recommendation 4: Bring the PDA into the consultation*

Many PDAs are given to patients as supplementary material to be used at home (8), with the benefit that the patient can go over the material at their own pace. One drawback is that it is difficult to determine whether the patient used the PDA, whether it was used correctly, and to what extent it had an impact on the decision process. There is evidence that simplified PDAs designed for use in the consultation itself stimulate dialogue and patient engagement (31). Developers could consider a combination of a 'regular' PDA for home use combined with a simpler version that clinicians can use in consultations. Alternatively, the regular PDA can be incorporated into the consultation in different formats. An electronic PDA could be connected to the hospital's electronic health record (EHR) system so that patient preferences or other relevant details are accessible to clinicians. Pure paper-based PDAs being used in consultations as part of an SDM process have also shown promising results (32). Randomized controlled trials show little difference in electronic vs. paper PDAs in terms of patient knowledge about prostate cancer, patient participation in decision-making, and decisional conflict, but find that PDAs combined with an SDM talk are more effective than PDAs given as a stand-alone tool (33).

*Recommendation 5: Obtain a certification mark*

Reliability and balanced information provided by the PDA was a priority for the clinicians we interviewed. As PDAs can potentially influence treatment choices there needs to be a way to maintain patient safety (34). Current criteria such as the International Patient Decision Aid Standards (IPDAS) outline a step-by-step process to guide PDA development to maximize quality, reliability, and user acceptance (35). This process is prescriptive however and may not necessarily guarantee a generally accepted measure of quality. Certification such as CE marking may support implementation by ensuring that the PDA meets well-established quality standards. This provides incentives for clinicians, hospital administrators, patients and insurers to use them in practice (36).

*Recommendation 6: Engage additional staff in PDA delivery*

With time constraints being a prominent barrier to SDM among our clinicians and other studies (13), there may be a role for additional personnel in the delivery and use of the PDA. Urologists interviewed for this study considered oncology nurses to be the ideal channel through which to offer the PDA to patients. They felt that nurses are better equipped to guide patients through the PDA if required and answer any questions. Our interviews with nurses, however, revealed that at present they are not trained to help patients make treatment decisions. Findings from other countries also show that nurses are not trained to specifically engage in SDM or introduce PDAs (37-39). Research highlights the importance of nurses receiving such training (40), particularly since SDM is not only relevant to major treatment decisions but also small decisions along the treatment trajectory where nurse support is invaluable (41).

*Recommendation 7: Training*

Clinicians who receive training in SDM report greater confidence in their abilities and increased positive attitudes towards SDM (42). PDAs are most effective when they are combined with an SDM conversation than when given to the patient as a stand-alone tool (33). Therefore, clinicians who are skilled in SDM are in a better position to deploy and use PDAs as part of their practice. While there is no consensus in the literature regarding the specific core competencies clinicians need in order to effectively involve patients in decision-making, prior findings highlight the importance of developing relational competencies (creating quality interactions between clinicians and patients) and risk communication competencies (helping patients understand treatment options, risks, and so on.) (43)

*Recommendation 8: Create appropriate incentives*

External incentives such as volume quotas for certain treatments pose another challenge to PDA implementation that has not been adequately addressed in the literature. Patients who are better informed about all relevant aspects of their disease and treatment tend towards choosing more conservative treatments (3). Dutch healthcare insurers are becoming more knowledgeable about SDM and supporting methods to ensure that patients receive treatments tailored to their profile and preferences (44). Insurers can stimulate patients and clinicians to take a more collaborative role in treatment decisions, for instance by paying clinicians for the use of PDAs or making their use a prerequisite for reimbursing certain procedures. Inclusion of a reimbursement code for the PDA can help standardize the process and speed up its usage, particularly since insurers see evidence of PDA use as one way to make the SDM process more visible and are heavily influencing hospitals to implement them (45). Moving from a volume-based to value-based healthcare model requires foremost collaboration between clinicians, hospital administrators, patients, patient advocacy groups to define appropriate outcome metrics and develop strategies to measure, analyze and disseminate the findings to continuously improve care quality (46).

*Recommendation 9: Measurement and evaluation*

Our clinicians believed that evidence of improved outcomes would provide a greater impetus to use and recommend PDAs to patients and take part in SDM. Currently, there are a variety of validated instruments to measure the effects of SDM along different dimensions. These include the OPTION scale measuring the level of SDM itself (47), the SDM-Q-9 measuring patient and clinician participation in the SDM process (48), the decisional conflict scale (49), and patients' post-treatment decision regret (50). In addition, routinely collected patient-reported outcomes (PROMs) may be useful in evaluating SDM initiatives or as a basis for values clarification (51). Care must be taken however when interpreting SDM outcome measures. In particular, what constitutes a 'good' clinical decision is open to debate (52) and outcomes may not be the most reliable indicator of decision quality. For instance, a patient's evaluation of their decision may change over time as long-term consequences become more apparent. More research is needed on methods to evaluate the decision-making process and its effects.

To our knowledge this is the first study that explores SDM and PDA implementation in parallel. We regard PDAs as a support to SDM and explore their implementation as a synergistic process. In taking this broader perspective, our study reveals factors that may affect PDA implementation that are not directly related to the PDA itself, for instance volume quotas. Another contribution of our work is the focus on clinician suggestions for implementation; buy-in from clinicians is crucial when introducing any innovation, and it is reasonable to assume that clinicians will be more open to strategies originating from their peers.

Despite our best efforts to remain objective, our study may suffer from bias associated with qualitative research, namely interpretation bias and sampling bias. To the extent possible we have tried to avoid the former by having participants review their interview transcripts and by coding in parallel with two researchers. Our sample size is small and limited to one clinic and therefore care should be taken when extrapolating our results to different contexts. Nevertheless, our findings largely correspond with existing literature, indicating that some level of generalizability is maintained. Also, although the context of our study was prostate cancer management, our recommendations may be generalizable to other preference-sensitive contexts in which a choice needs to be made between two or more comparable treatment options. Future work might consider the role of oncology nurses in SDM/PDA implementation, as this group was underrepresented in our study. It may also be valuable to conduct a similar study from the patient perspective. Future research might also explore how the activities and communication methods of patient organizations can be leveraged in SDM and PDA implementation, as well as those of healthcare insurers.

**Conclusion**

PDAs are a useful supporting tool for SDM by helping patients clarify their preferences and engage in the decision-making process. Implementation is a multifaceted process requiring action on several levels. Barriers and facilitators exist at the individual patient-clinician level, at the organizational level, and in the external environment. A PDA implementation strategy should include an assessment of the extent to which the implementation environment is conducive to SDM. With this in mind, coordinated actions can be taken at the individual, administrative, and policy level to support patient engagement.

**List of abbreviations**

PDA – Patient decision aid

SDM – Shared decision-making

IPDAS – International Patient Decision Aid Standards